\documentclass[aps,showpacs,twocolumn,superscriptaddress,prb,floatfix, 10pt, longbibliography]{revtex4-1}
\usepackage{graphicx}
\usepackage{subfigure}
\usepackage{amssymb,amsmath,cancel}
\usepackage{comment}
\usepackage{xcolor}
\usepackage{array}

\begin{document}

\title{Topological order in spin nematics from the quantum melting of a disclination lattice}

\author{Predrag Nikoli\'c}
\affiliation{Department of Physics and Astronomy,\\George Mason University, Fairfax, VA 22030, USA}
\affiliation{Institute for Quantum Matter at Johns Hopkins University, Baltimore, MD 21218, USA}

\date{\today}

\begin{abstract}

The topological defects of Spin($n+1$) nematics in two spatial dimensions, known as disclinations, are characterized by the $\pi_1(\mathbb{R}P^n) = \textrm{Z}_2$ homotopy group for $n\ge2$. We argue that incompressible quantum liquids of disclinations can exist as stable low-temperature phases and host composite quasiparticles which combine a fractional amount of fundamental Z$_2$ charge with a unit of topological charge. The four-fold topological ground state degeneracy on a torus admits a fermionic or semionic quasiparticle exchange statistics. The topological non-triviality of these states is visible in the existence of protected gapless edge modes. While the fermionic nematic and gapped Z$_2$ spin liquids have equivalent topological orders, they are still thermodynamically distinct due to having different edge modes, in analogy to the topologically non-trivial and trivial states of quantum spin-Hall systems. The analysis proceeds by recasting the Z$_2$ gauge theory of spin nematics as a continuum limit theory with a larger gauge structure. Nematic fractionalization parallels that of a quantum Hall liquid, but the large gauge symmetry restores the time-reversal symmetry and restricts the quasiparticle fusion rules and statistics. The conclusions from field theory analysis are complemented with the construction of plausible host microscopic models.

\end{abstract}

\maketitle

\section{Introduction}

Incompressible quantum liquids are gapped phases of matter without a broken symmetry, but with unconventional properties tied to a non-trivial topology or strong correlations. The most studied examples are quantum Hall states \cite{WenQFT2004} and gapped spin liquids \cite{anderson73b, anderson87c, Kitaev2006b, Savary2016}. A large class of such states can be described in the continuum limit as a fluid of abundant delocalized topological defects characterized by $\pi_n(S^n)$ and $\pi_{2n-1}(S^n)$ homotopy invariants \cite{Nikolic2019, Nikolic2023b}. If a parent Higgs phase with long range order hosts a lattice of quantized topological defects, then its first-order melting transition can produce a liquid of topological defects whose topological charge, a homotopy invariant, is still quantized and conserved by the surviving field coherence across inter-defect distances -- a notion made precise via instanton confinement in the renormalization group \cite{Nikolic2023a}. Numerical studies have indeed observed fractional quantum Hall liquids arising from the quantum melting of Abrikosov vortex lattices in bosonic superfluids \cite{Cooper2001}. The resulting simultaneous conservation of fundamental and topological charge necessarily binds the two charges into composite quasiparticles with some form of fractionalization. Topological order, or ground state degeneracy without a symmetry breaking on topological manifolds, is then a direct consequence of the fractional quasiparticle braiding statistics \cite{WenQFT2004}.

Here we explore the incompressible quantum liquids of topological defects in two spatial dimensions characterized by the $\pi_1(\mathbb{R}P^n)$ homotopy invariant. Nematic order parameters which break an $\textrm{O}(n+1)/\textrm{Z}_2$ rotational symmetry admit such topological defects. If the nematic degrees of freedom can rotate in an $n+1$ dimensional internal space, then the defects are equivalent to vortices for $n=1$, or known as disclinations if $n\ge2$. The latter is of interest in this study. Only two topologically distinct equivalence classes are available to disclinations, so their topological charge group is just Z$_2$. The fundamental charge transported by the Higgs modes of the nematic field is also Z$_2$-like according to the Dirac ``monopole'' quantization principle. We will show that a quantum liquid of disclinations has much in common with fractional quantum Hall states, but also notable differences attributed to their Z$_2$ instead of the full $\mathbb{Z}$ charge algebra.

Nematic fluids have been studied in many different contexts. They are of great interest in statistical mechanics and soft condensed matter \cite{Lebwohl1972, Mermin1979, Michel1980, deGennes1995, Toner1993, Toner1995a, Toner1995b, Gonnella1995, Zaanen2015}, and especially in strongly correlated quantum materials \cite{Pomeranchuk1958, Levy1971, Coleman1991, Vojta2009, Mackenzie2010, Cvetkovic2006, Cvetkovic2007, Metlitski2010, Momoi2013, Sachdev2016, Sachdev2019, Grover2010, Zaanen2004a, Hughes2013, Hughes2015, Kaul2015, Nevidomskyy2019b, Harada2007, Grover2007, Grover2011, Podolsky2005, Kawashima2002, Tsunetsugu2006, Kolezhuk2003} and cold atom gases \cite{Demler2002b, Vishwanath2009, Pixley2018, Yip2003, Imambekov2003, Ueda2013}. Stripe (smectic) and nematic states have been explored as candidates for the pseudogap state and competing orders in cuprate high-$T_c$ superconductors \cite{Zaanen1989, Rice1989, Schulz1989, Machida1989, kivelson98, emery99b, zhang02, kivelson03, CastroNeto2003, Vojta2010}, with support from numerous experimental observations of anisotropic transport and charge/spin fluctuations \cite{Tranquada1995, Tranquada1996, Lavrov2002, Keimer2008, Lawler2010, Mesaros2011}. Similar anisotropies have also been discussed in quantum Hall systems \cite{fradkin99a, Lilly99p08178, Lilly99p824, Stormer1999, Abanin2010, Mulligan2011, Barkeshli2012, Maciejko2013, Fradkin2014, Fradkin2016, Sondhi2017}, iron-based superconductors \cite{Fang2008, Xu2008a, Davis2010,  Davis2010, Analytis2012a, Schmalian2012, Fernandez2014, Wang2015c, Schmalian2015, Berg2016}, and other materials \cite{Oganesyan2007, Raghu2009, Sun2009, Oppeneer2011, Fujimoto2011, Galitski2015, Herbut2015, Herbut2016b, DasSarma2023}. In all these cases, the stripe and nematic degrees of freedom are complex structures built from microscopic electron spins and charge, including sometimes lattice distortions; their internal rotation space is confined to that of the typically two-dimensional lattice, and it is discrete. While this physics serves as a partial motivation for the present study, we will be focused on isotropic nematics with a continuous and higher-dimensional rotation freedom, because we wish to confront the qualitatively distinctive dynamics of elementary Z$_2$ topological defects in the continuum limit.

Spin nematics are classic systems whose local degrees of freedom have an orientation without direction in a continuous isotropic internal space \cite{Toner1995a, Toner1995b, Toner1993, Gonnella1995}. The nature of the disordered phase in quantum spin nematic models on a lattice has been analyzed in various contexts \cite{Coleman1991, zhang02, Kawashima2002, Kolezhuk2003, Podolsky2005, Tsunetsugu2006, Harada2007, Grover2007, Grover2011, Kaul2015, Nevidomskyy2019b}, seeking realizations of quantum spin liquids and taking inspiration from cuprates and frustrated magnets. In these studies, the loss nematic long range order can be viewed by duality as the ``condensation'' of Z$_2$ topological defects. The topological charge, or the number of condensed defects, is not a locally well-defined conserved quantity in such a condensate. Nevertheless, the ``odd'' Z$_2$ gauge theory that captures the defect condensate predicts topological order in this disordered phase as a result of geometric frustration which encodes the effect of a quantum Berry phase.

In this paper, we seek a fundamentally different route to a quantum disordered phase. We envision a parent Higgs phase with a lattice of topological defects, and ask what happens when this defect lattice melts by quantum fluctuations. In this regime, the correlation length $\xi$ of the nematic field becomes suddenly limited by the ``magnetic length'', i.e. the mean distance between the defects, which we assume to be significantly larger than the ultra-violet cut-off length or the microscopic lattice constant $\delta a$. If the confinement length $\lambda$ of instantons, which create or annihilate individual defects, is smaller than $\xi$, then the nematic field is coherent across sufficiently large length and time scales to protect the conservation and quantization of topological charge. This is sharply characterized with a generalized Wilson loop operator, and established by a renormalization group of the instanton gas \cite{Nikolic2023a}. The resulting incompressible quantum liquid is a nematic analogue of a fractional quantum Hall state. The dynamics of this state can be still captured with a continuum limit theory involving the original nematic field, as long as a Z$_2$ gauge field is introduced to keep track of the delocalized defects. Additional ``large'' gauge symmetry turns out to be necessary in the continuum limit as well, but it helps to identify the ``gauge-invariant'' observable effects of fractionalization. We find that topological order with fermion or semion fractional quasiparticles arises in this microscopically bosonic system.

Our approach provides a view of the Z$_2$ fractionalization without the usually employed slave boson method, and in the regime $\xi\gg\delta a$ different than the usually studied $\xi\approx \delta a$. The origin of the Z$_2$ fractionalization is found in the Berry curvature of disclinations, which shapes a ``doubled'' Chern-Simons coupling in the continuum-limit low-energy effective theory. Similar Chern-Simons theories have been proposed before in the context of Z$_2$ spin liquids \cite{Kou2008, Xu2009b}, but only to provide a mutual semionic statistics between fractionalized ``charges'' and vortices. Here, the Chern-Simons coupling also fractionalizes the ``charge'' with a degree of predictive rigor, and relieves the slave boson method of its duty to introduce fractionalization phenomenologically. The phase with fermionic quasiparticles that we find is closely related to the gapped RVB spin liquid, but differs from it by its gapless edge modes \cite{Kou2008}, resembling a quantum Hall liquid or more accurately a topologically non-trivial spin-Hall liquid with time-reversal symmetry.

The rest of the paper is organized as follows. We begin by constructing a continuum limit Z$_2$ gauge theory of nematic fluids in Section \ref{secFieldTheory}, focusing on the conserved currents in Section \ref{secCurrents}. The incompressible quantum liquids of disclinations are discussed in Section \ref{secFract}. Here we derive the fusion rules for the fractional quasiparticles in such liquids, and analyze their exchange statistics in relation to the topological ground state degeneracy. All conclusions and a discussion of some further implications are summarized in Section \ref{secConclusions}. Some technical details are given in the appendices. Specifically, Appendix \ref{appModel} presents model lattice Hamiltonians which are expected to host the envisioned nematic quantum liquids in their phase diagrams. 

We use units $\hbar=c=1$ throughout the paper, as well as Einstein's convention for the summation over repeated indices. Lower Greek indices $\mu,\nu,\dots$ are space-time directions and lower Latin indices $i,j,\dots$ are spatial directions. Upper indices $a,b,\dots$ denote internal spin-projection directions. We work in imaginary time (without distinguishing between upper and lower space-time indices), except when discussing the field equations of motion.

\bigskip

\section{Field theory of a spin nematic}\label{secFieldTheory}

Nematic degrees of freedom have spatial orientation without direction. They can be represented by a unit vector $\hat{\bf d}$, known as director, whose states $\hat{\bf d} \equiv -\hat{\bf d}$ are identified. A non-redundant representation is the symmetric tensor $D^{ab} = \hat{d}^a\hat{d}^b$, i.e. a projection matrix which reproduces $\hat{\bf d}$ as its eigenvector corresponding to the unique non-zero eigenvalue. Another widely used representation is the traceless tensor $S^{ab} = D^{ab} - \delta^{ab}/\textrm{tr}(1)$. We will assume that $\hat{\bf d}$ originates in the microscopic or effective $S\ge 1$ spins of some particles. This can in principle produce a spin nematic with isotropic dynamics at low energies if the Hamiltonian has appropriate symmetries.

The simplest theory of an isotropic nematic is given by the imaginary-time Lagrangian density
\begin{equation}\label{FieldTheory1}
\mathcal{L} = \frac{K}{2}(\partial_{\mu}D^{ab}+2a_{0}\delta_{\mu,0}D^{ab})^{2} + \mathcal{L}_D \ .
\end{equation}
This is a relativistic theory when the ``chemical potential'' $a_0$ is zero. The spin Berry phase is absent at least because the local symmetry under $\hat{\bf d} \to -\hat{\bf d}$, to be gauged shortly, prohibits ferromagnetism according to Elitzur's theorem \cite{Elitzur1975}. We will be interested in the phases where the nematic field retains a meaningful finite magnitude across length scales much larger than the ultra-violet cut-off of the theory, i.e. a lattice constant. Then, normalizing $\hat{\bf d}$ imparts a unit-trace on the nematic tensor $D^{ab}$, as well as its square. If the theory is meant to access a second order transition between ordered and isotropic phases, then the fields can be softened and
\begin{equation}
\mathcal{L}_D = u (D^{aa}-1)^2 + v (D^{ab}D^{ab}-1)^2 + w (D^{ab}-A^{ac}A^{bc})^2 \nonumber
\end{equation}
can be added to the Lagrangian density, where $A^{ab}$ is an auxiliary real matrix randomly sampled in the path integral; $D^{ab}$ is coerced to a projection matrix when $u,v,w>0$ are large.

\begin{figure}[!t]
\subfigure[{}]{\includegraphics[height=0.8in]{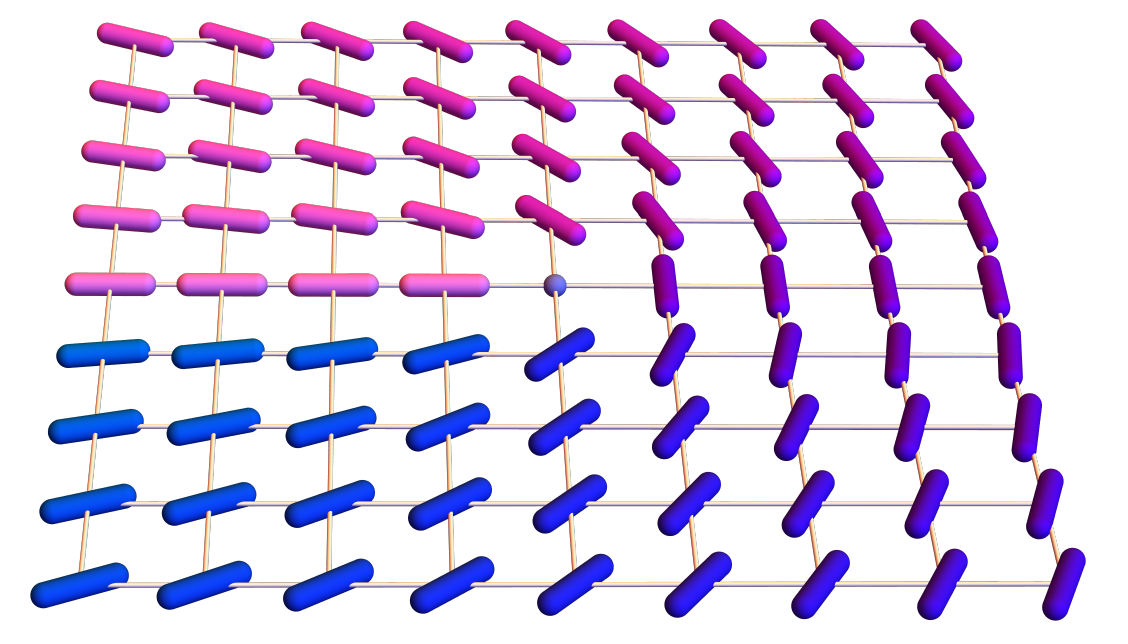}}\hspace{0.1in}
\subfigure[{}]{\includegraphics[height=0.8in]{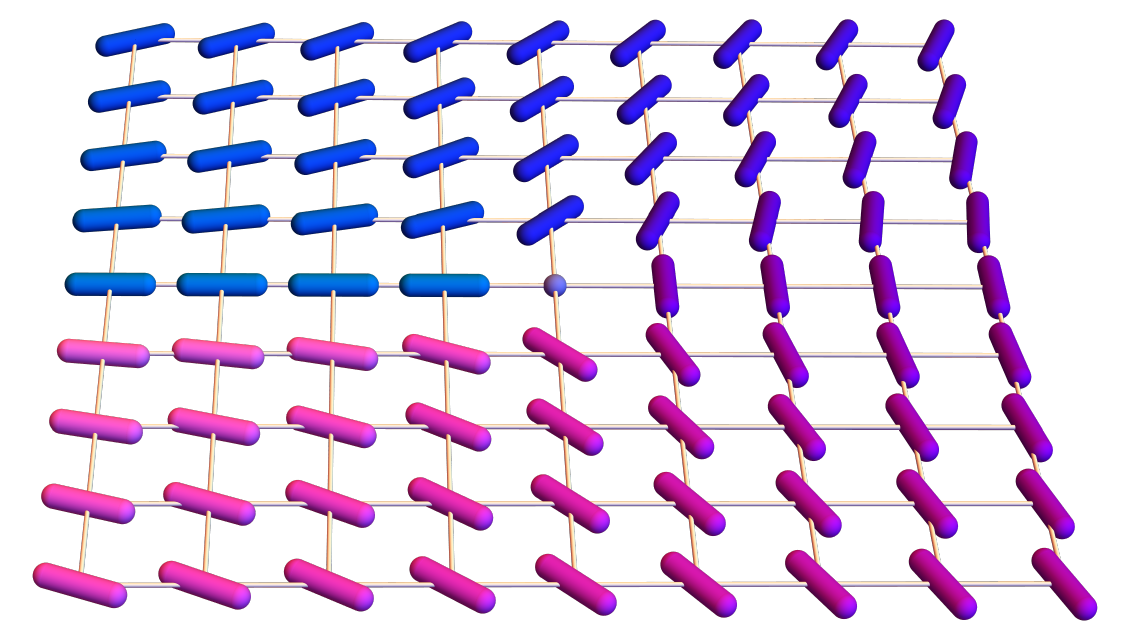}}
\caption{\label{Disclinations}Disclinations in the nematic order. If nematic order parameter can rotate only in a two-dimensional $N=2$ plane, then (a) and (b) are topologically distinct and can be regarded as vortices of opposite topological charge. If the order parameter can rotate in a higher-dimensional $N\ge3$ space, then (a) and (b) are topologically equivalent and can be smoothly deformed into one another. Note $N=n+1$.}
\end{figure}

\begin{figure}[!t]
\subfigure[{}]{\includegraphics[height=0.7in]{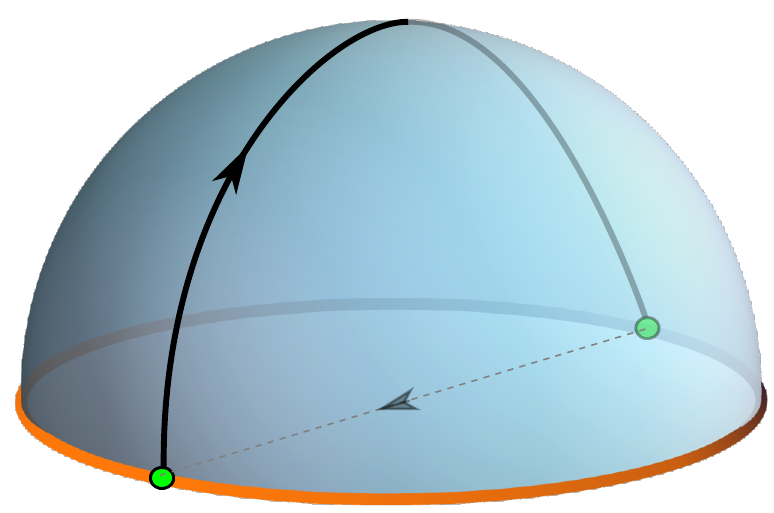}}\hspace{0.05in}
\subfigure[{}]{\includegraphics[height=0.7in]{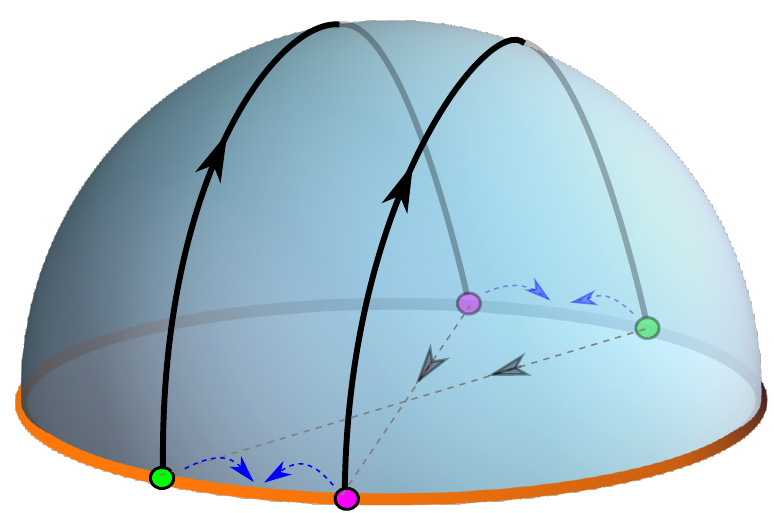}}\hspace{0.05in}
\subfigure[{}]{\includegraphics[height=0.7in]{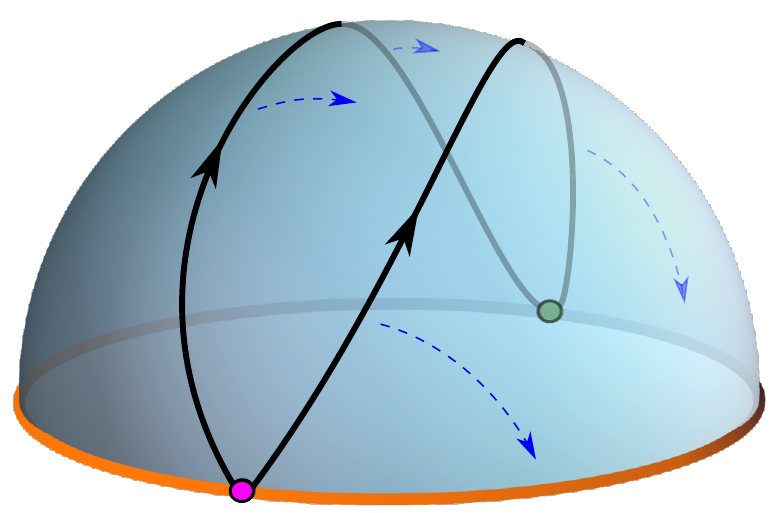}}
\subfigure[{}]{\includegraphics[height=0.7in]{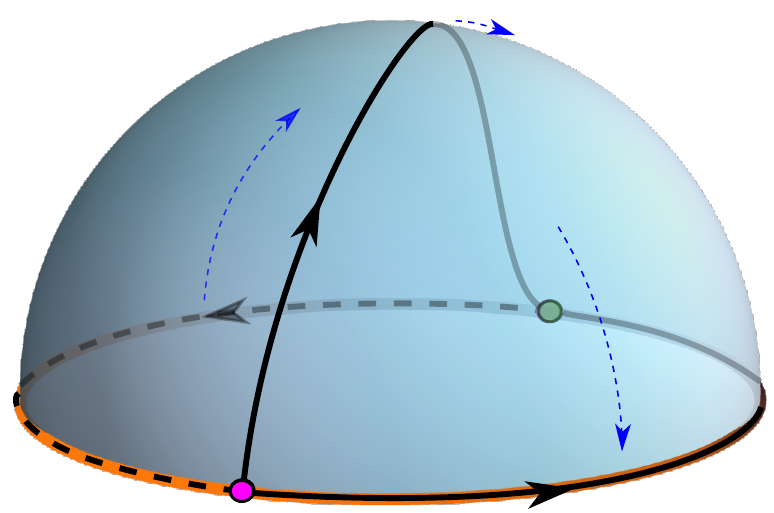}}\hspace{0.05in}
\subfigure[{}]{\includegraphics[height=0.7in]{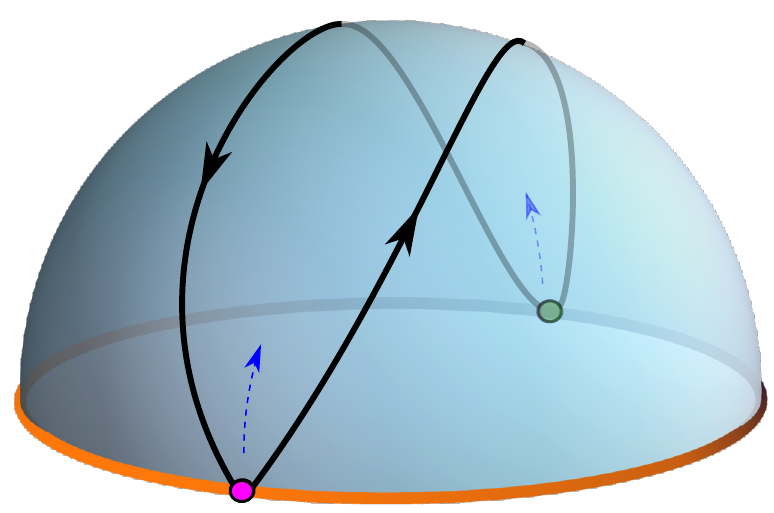}}\hspace{0.05in}
\subfigure[{}]{\includegraphics[height=0.7in]{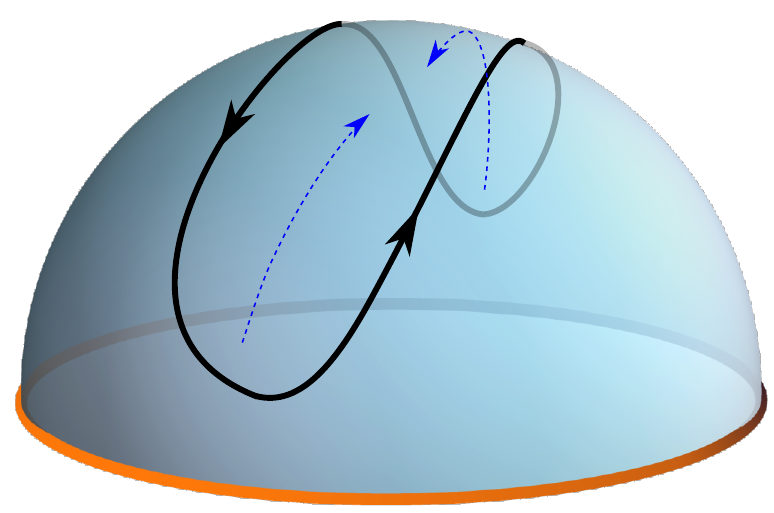}}
\caption{\label{LoopContracting}An $\mathbb{R}P^2$ nematic order parameter lives on a sphere whose antipodal points are identified ($\hat{\bf d}\equiv-\hat{\bf d}$), or equivalently on a hemisphere with a ring boundary whose antipodal points are identified. (a) A non-contractible loop. The path with an arrow represents the loop, and the points on the hemisphere represent $\hat{\bf d}$. This configuration depicts the variations of $\hat{\bf d}$ along a closed lattice path around the disclination singularity in Fig.\ref{Disclinations}(a). Reshaping the loop on the hemisphere smoothly changes the variations of $\hat{\bf d}$, but its two antipodal points on the ring can never be made to meet. (b)-(f) A contractible loop, corresponding to the difference between the two patterns shown in Fig.\ref{Disclinations}(a,b) for $n=2$ ($N=3$). Blue dashed arrows indicate the directions in which loop segments can be moved to contract it.}
\end{figure}

The nematic order parameter obtained from $\hat{\bf d} \in S^n$ ($n$-sphere) spans the ``projective plane'' manifold $\mathbb{R}P^n$. The non-trivial homotopy groups $\pi_1(\mathbb{R}P^n)=\textrm{Z}_2$ for $n\ge2$ imply that the nematic field has singular topological defects in $d=2$ spatial dimensions whose protected ``charge'' takes only two topologically distinct values. These defects are known as disclinations \cite{Toner1993, Mermin1979} and explained in Fig.~\ref{Disclinations} and Fig.~\ref{LoopContracting}. Nematic fields can support other topological defects in various circumstances owing to $\pi_1(\mathbb{R}P^1)=\mathbb{Z}$ (vortices in $d=2$), and $\pi_k(\mathbb{R}P^n)=\pi_k(S^n)$ for $k>1$ (hedgehogs, hopfions in $d=3$). These will not be analyzed here since their dynamics must have much in common with their counterparts made from vector fields \cite{Nikolic2019, Nikolic2023b}.

The proliferation of topological defects is detrimental to long-range order. Our goal is to characterize the quantum disordered phases of spin nematics shaped by mobile disclinations whose Z$_2$ topological charge remains conserved. The nematic field must be coherent at sufficiently long length scales $\xi$ in order to preserve the quantization and conservation of the Z$_2$ topological charge. This requirement is thermodynamically characterized \cite{Nikolic2023a} as the confinement of instantons to distances $\lambda<\xi$. Quantum Hall liquids are examples of such states, featuring abundant, mobile and conserved $\pi_1(S^1)$ topological defects. It is indeed possible to enter such a state from an ordered phase by quantum melting of a defect (Abrikosov) lattice \cite{Cooper2001}. Here we seek a continuum limit description for the $\pi_1(\mathbb{R}P^n)=\textrm{Z}_2$ disclination lattice melting, which must exist when $\xi$ is much larger than the microscopic lattice constant and the instantons are confined.

\begin{figure}
\subfigure[{}]{\includegraphics[height=1.3in]{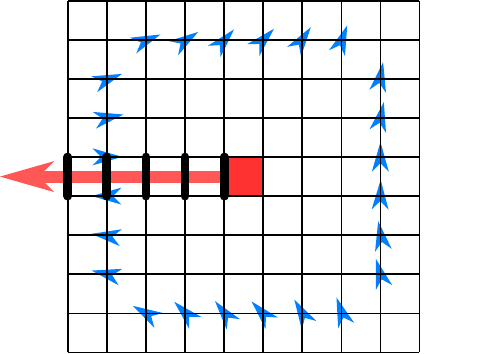}}
\subfigure[{}]{\includegraphics[height=1.3in]{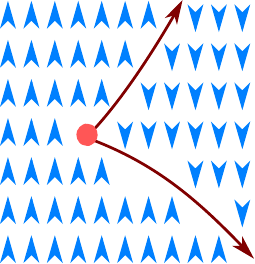}}
\caption{\label{Vison}(a) The lattice Z$_2$ gauge field in the vicinity of a disclination: $\sigma_{ij}=-1$ is indicated with thick links, otherwise $\sigma_{ij}=+1$. The colored square holds the singularity of the disclination, which is a vison of the Z$_2$ gauge field, and the red arrow shows the ``Dirac string'' attached to it. The pattern of the director field $\hat{\bf d}$ on one loop around the singularity is shown with blue arrows; compare with Fig.\ref{Disclinations}(b). (b) A hypothetical singularity with two outgoing strings and trivial Z$_2$ topological charge.}
\end{figure}

A general method to capture the dynamics of topological defects involves gauging the field theory. Here we want to introduce a Z$_2$ gauge field which can describe the disclinations of an $n\ge2$ nematic. The following imaginary-time action on a square lattice with sites $i$ achieves this goal \cite{Toner1993, senthil00}
\begin{equation}\label{LatticeTheory1}
S = -\kappa\sum_{j\in i}\hat{\bf d}_{i}\sigma_{ij}\hat{\bf d}_{j}-\gamma\sum_{\square}\prod_{ij}^{\square}\sigma_{ij} + S_B \ .
\end{equation}
The Z$_2$ gauge field $\sigma_{ij}\in\lbrace\pm1\rbrace$ lives on the space-time lattice links $ij$. A Berry phase term $S_B$ is non-trivial for odd spin representations $\lbrace \frac{1}{2}, \frac{3}{2}, \dots\rbrace$, as shown in Appendix \ref{appModel}. One can see in Fig.\ref{Disclinations} that an otherwise smooth $\hat{\bf d}$ must change sign within a spatial plane ($\tau=\textrm{const.}$) across a semi-infinite string that emanates from the defect singularity. The gauge field absorbs this sign change by $\sigma_{ij}=-1$ on all lattice links that intersect the string, keeping the ``exchange'' energy low (see Fig.\ref{Vison}). The $\gamma>0$ term is the defect core energy. The orientation and shape of strings are arbitrary and affected only by local Z$_2$ gauge transformations:
\begin{equation}
\hat{\bf d}_i \to -\hat{\bf d}_i \quad,\quad (\forall j\in i)\quad \sigma_{ij}\to-\sigma_{ij} \ ,
\end{equation}
where $j\in i$ denotes all nearest neighbor sites $j$ of the site $i$. Integrating out $\sigma_{ij}$ preserves the Z$_2$ gauge symmetry and thus couples the $\hat{\bf d}_i \hat{\bf d}_j$ scalars only on the links of small closed loops, giving (\ref{FieldTheory1}) in the continuum limit.

Now, how could a Z$_2$ gauge theory be formulated in the continuum limit without losing the discrete and cyclic nature of the gauge field? Consider the continuum-limit space-time configurations $\hat{\bf d}({\bf x})$ that correspond to well-defined disclinations on a lattice. The gauge field is a scalar under internal rotations and a vector under space-time rotations. The following quantity of this kind can indeed capture the sign change of $\hat{\bf d}$ at a fixed time $\tau$ across a string in the continuum limit,
\begin{equation}\label{String1}
\hat{d}^a({\bf x})\partial_i\hat{d}^a({\bf x}) = -2 \epsilon_{ij} \int dX_j \,\delta ({\bf x}-{\bf X}) \ .
\end{equation}
We assumed $|\hat{\bf d}|=1$ and expressed the points on the string at time $\tau$ with ${\bf X}(u)$, parametrized by $u\ge0$ ($\epsilon_{ij}\equiv\epsilon_{0ij}$ is the Levi-Civita tensor). But, the expression on the left-hand side has an ambiguous sign at the $\hat{\bf d}\to-\hat{\bf d}$ discontinuity, so we hereby attempt to regularize it by giving each string an orientation via the direction of $d{\bf X}$. We will examine the consistency of this regularization later. The gauge field which compensates the discontinuity must be the vector $\sigma_i$ determined by the right-hand side. The resulting Dirac string indeed obtains if one zooms-out the configuration in Fig.\ref{Vison}. As long as the gauge field is so singular, we can read out its flux $\phi$ through a loop $\mathcal{C}$ and compute from it the Z$_2$ topological charge $\mathcal{N}$ inside the loop:
\begin{equation}\label{Z2flux}
\mathcal{N} = (-1)^\phi \quad,\quad \phi = \frac{1}{2}\oint\limits_{\mathcal{C}} dx_i \sigma_i 
\end{equation}
This is made possible because $\phi$ registers $\pm 1$ each time $\mathcal{C}$ crosses a string; all possible ambiguities associated with the mentioned regularization do not matter for the topological charge $\mathcal{N}\in\lbrace{\pm 1}\rbrace$, and $\mathcal{N}=-1$ corresponds to a disclination or vison. It should be noted that an alternative field theory could interpret the Z$_2$ gauge field as a scalar labeled by the spatial direction $i$. This is effectively done in the lattice theory (\ref{LatticeTheory1}), but would make the continuum limit anisotropic.

The challenge for the continuum limit is to capture the disclination core energy as a function of the Z$_2$-quantized flux. This can be faithfully accomplished only with a compact gauge theory, which in the continuum limit necessarily contains an explicit ultra-violet cut-off length scale $\delta a$ and reduces the continuous rotational symmetry to a discrete one:
\begin{equation}\label{FieldTheory2}
\mathcal{L} = -\gamma_c\sum_\mu\cos (\delta a^2 \epsilon_{\mu\nu\lambda}\partial_{\nu}\sigma_{\lambda}) + \cdots \ .
\end{equation}
Aside from the undesirable lack of symmetry, this theory, if properly tuned, admits large gauge transformations in which the quantized singular flux $\phi$ changes by 2; this implements the Z$_2$ algebra. For example, the configuration in Fig.\ref{Vison}(b) is just a large gauge transformation of the uniform ordered state.

The low-energy effective theory with all desired symmetries must be non-compact,
\begin{equation}\label{FieldTheory3}
\mathcal{L} = \frac{\kappa}{2}(\partial_{\mu}\hat{d}^{a}+c\,\sigma_{\mu}\hat{d}^{a}+a_{0}\delta_{\mu,0}\hat{d}^{a})^{2}+\frac{1}{2e^{2}}(\epsilon_{\mu\nu\lambda}\partial_{\nu}\sigma_{\lambda})^{2} + \dots
\end{equation}
The cut-off $\delta a$ is absorbed into the coupling constants $\kappa$, $e^2$, and the ellipsis represents the terms needed to keep $|\hat{\bf d}|\to1$. The fundamental Z$_2$ charge unit is $c$. The only local symmetry is Z$_2$; the gauge field has a deceptive U(1) form and Maxwell term, but the non-compact gradient term still breaks the U(1) gauge symmetry down to Z$_2$. The main issue here is that the non-compact Maxwell term does not cyclically constrain the quantization of topological charge. This is resolved by pushing the $|\phi|>1$ topological charges to higher energies and maintaining incompressibility for the $\phi=\pm1$ disclinations. Importantly, every nematic field configuration can be represented in the low-energy sector with the following procedure. Identify disclination singularities and decide how to attach to them some oriented strings for the required sign changes of $\hat{\bf d}$. Define the ``charge'' $\phi$ of each singularity as the number of its outgoing minus the number of incoming strings. Then, reduce every singular charge to $0$ or $\pm 1$ by altering the orientations of some strings. Next, eliminate closed oriented loops of strings in any order until there are none left. Some loops may include semi-infinite strings that terminate at ``hidden'' far-away singularities. What remains is a disjoint graph of $\phi=\pm1$ singularity pairs connected by one string each. The final state minimizes the non-compact core energy of disclinations, but still has a ``large'' gauge redundancy associated with an arbitrary $\phi=\pm1$ ``charge'' assignments to disclinations.

In order to understand better the full set of gauge transformations, let us consider the 2+1D world view of the path integral. The low-energy $\phi=\pm1$ vortices of the $\sigma_{\mu}$ gauge field form worldlines embedded in the 2+1D space-time, which are oriented loops of arbitrary shape and size. The director field $\hat{d}^{a}$ changes sign across arbitrary sheets or membranes bounded by the vortex worldlines, and $\sigma_{\mu}$ must screen out the gradient energy of these discontinuities. Local or small Z$_2$ gauge transformations are smooth changes of the sheet shapes, while large gauge transformations are reorientations of the vortex worldlines. The latter restore the $Z_{2}$ algebra for disclinations from the U(1) algebra of vortices. Now one has the freedom to choose a finite space-time region whose boundary does not intersect any worldline or sheet and transform $\sigma_{\mu}({\bf x},\tau)\to-\sigma_{\mu}({\bf x},\tau)$ at all points inside the region. But then we must also give the Z$_2$ charge unit a space-time dependence and change its sign $c({\bf x},\tau)\to-c({\bf x},\tau)$ inside the region. This is instrumental for preserving the Z$_2$ charge current
\begin{equation}\label{Z2current}
j_\mu = \hat{d}^a \partial_\mu \hat{d}^a + c\,\sigma_\mu + a_0 \delta_{\mu,0}
\end{equation}
under gauge transformations, and faithful to the spirit of representing the Z$_2$ charge with a relativistic U(1)-like counterpart which couples to the U(1)-like gauge field. Importantly, the large gauge transformation preserves the homotopy invariant given by the product of ``electric'' and ``magnetic'' charge units. The freedom to introduce space-time variations of $c=\pm1$ ultimately comes from the arbitrariness of singular director field gradients. We generalize (\ref{String1}) to
\begin{equation}\label{String2}
\hat{d}^{a}({\bf x})\partial_{\mu}\hat{d}^{a}({\bf x})=-2c({\bf x})\int dX_{\nu}dX_{\lambda}\epsilon_{\mu\nu\lambda}\,\delta({\bf x}-{\bf X})
\end{equation}
in space-time, where ${\bf X}(u,v)=(X_{0},X_{1},X_{2})$ are the points on the oriented sheet of $\hat{d}^a$ discontinuity, and $c({\bf x})$ provides a local interpretation of the regularization. In fact, $c({\bf x})$ is the gauge field that keeps track of ``large'' gauge transformations.

The simplest way to introduce incompressible disclinations into the nematic state is to modify the non-compact Maxwell term of (\ref{FieldTheory3}) into
\begin{equation}\label{MagnField1}
\mathcal{L}_{\textrm{M}} = \frac{1}{4e^{4}}(\boldsymbol{\nabla}\times\boldsymbol{\sigma}-B\hat{{\bf z}})^{2}(\boldsymbol{\nabla}\times\boldsymbol{\sigma}+B\hat{{\bf z}})^{2} \ .
\end{equation}
As long as the gradient Lagrangian term quantizes the flux, the external ``magnetic'' field $B$ stimulates a finite density of $\phi=\pm1$ flux quanta. The resulting disclinations must be localized and form a lattice in the Higgs phase. Even though the average flux can be zero, singularities cannot massively annihilate without an energy penalty. In fact, $\mathcal{L}_{\textrm{M}}$ includes repulsive interactions between disclinations which prevents their annihilation and the formation of higher-energy $|\phi|>1$ singularities. Defect incompressibility survives in the incompressible quantum liquid.

\subsection{Currents}\label{secCurrents}

The non-compact theory (\ref{FieldTheory3}) represents the Z$_2$ topological charge of disclinations as a quantized $\phi=\pm1$ vorticity of the vector gauge field $\sigma_\mu$. We define the current density $j_\mu$ of the conjugate or ``fundamental'' Z$_2$ charge from the conservation law $\partial\mathcal{L}/\partial\sigma_\mu\to0$,
\begin{equation}\label{Z2current}
j_\mu = \partial_\nu F_{\mu\nu} = \hat{d}^{a}\partial_{\mu}\hat{d}^{a}+c\,\sigma_{\mu}+a_{0}\delta_{\mu,0} \ ,
\end{equation}
where $F_{\mu\nu} = \partial_\mu\sigma_\nu-\partial_\nu\sigma_\mu$. The background Z$_2$ charge density is given by $j_0 = a_0$. In the Higgs phase, $|\hat{\bf d}|=1$ suppresses all Z$_2$ charge fluctuations beyond the background charge density. In fact, Z$_2$ charge currents are generally gapped by the virtue of being longitudinal modes of the director field. We can also construct the Z$_2$ charge current in terms of the nematic tensor by picking the same variations of the (\ref{FieldTheory1}) Lagrangian density and comparing to the above via $D^{ab}=\hat{d}^a\hat{d}^b$,
\begin{equation}\label{Z2current2}
J_{\mu} = D^{ab}(\partial_{\mu}+2a_0\delta_{\mu,0})D^{ab} = 2j_{\mu} \ .
\end{equation}
This explicitly shows that the tensor $D^{ab}$ carries twice the Z$_2$ charge of the director field modes, which amounts to zero upon charge quantization. Consequently, the tensor field cannot be coupled to the Z$_2$ gauge field.

Non-compact gauge theories implement charge quantization topologically. First, the matter field must support topological defects quantized by the invariant of some homotopy group. Since the gauge field flux can be associated with topological defects, the flux is quantized as an integer multiple of a flux quantum $m$ in appropriate units. Then, the minimal coupling between the matter and gauge fields introduces a fundamental charge unit $e$. Both $m$ and $e$ can be tuned by the rescaling of fields, i.e. the change of units, but the product $e\cdot m$ is fixed by the homotopy of the topological defects. This generalizes Dirac's monopole quantization. Microscopic excitations can carry fundamental charge $n_ee$ and topological charge $n_mm$, where $n_e,n_m\in\mathbb{Z}$. The homotopy $\pi_1(\mathbb{R}P^n)=\textrm{Z}_2$ for $n\ge2$ ensures that only two values of $n_e\cdot n_m\in\lbrace 0,1 \rbrace$ are topologically inequivalent, so that both the fundamental $n_e$ and topological charge $n_m$ must obey the Z$_2$ algebra ($1+1=0$).

The quantum nematic theory with $N$-dimensional internal space also possesses rotational symmetry, specifically $\textrm{O}(N)/\textrm{Z}_2$ because $\pi$-rotations are Z$_2$ gauge transformations. Therefore, a spin current should be conserved as well. This current carries $N-2$ spin indices. For the $N=3$ nematics of the physical spin, rotations of the vector and tensor fields
\begin{eqnarray}
\hat{d}^a &\to& \mathcal{R}^{aa'}\hat{d}^{a'} \\
D^{ab} &\to& \mathcal{R}^{aa'}\mathcal{R}^{bb'}D^{a'b'} \nonumber
\end{eqnarray}
by angle $\theta$ about axis $\hat{\bf n}$ are generated by the rotation operator
\begin{eqnarray}
\mathcal{R}^{ab}(\hat{{\bf n}},\theta) &=& \hat{n}^{a}\hat{n}^{b}+(\delta^{ab}-\hat{n}^{a}\hat{n}^{b})\cos\theta-\epsilon^{abc}\hat{n}^{c}\sin\theta \nonumber \\
&& \xrightarrow{\theta=d\theta\to 0}\delta^{ab}-\epsilon^{abc}\hat{n}^{c}d\theta \ .
\end{eqnarray}
The Lagrangian density is strictly invariant under global rotations, so the spin current derived using Noether's theorem is
\begin{equation}
J_{\mu}^{a}=\epsilon^{abc}D^{bp}\partial_{\mu}^{\phantom{x}}D^{cp} = \epsilon^{abc}\hat{d}^{b}\partial_{\mu}^{\phantom{x}}\hat{d}^{c} \ .
\end{equation}
Here we again utilized $|\hat{\bf d}|=1$. Note that a factor of 2 is absent in the conversion from $D^{ab}$ to $\hat{d}^a$.

\section{Topological order}\label{secFract}

Here we explore the possibility of stabilizing incompressible quantum liquids with topological order in spin nematics. We are interested in scenarios where a lattice of disclinations within an ordered nematic phase melts due to quantum fluctuations. The resulting liquid state can still conserve the quantized Z$_2$ topological charge of disclinations by the confinement of instantons \cite{Nikolic2023a}. Since the loss of long-range order also facilitates a local conservation of the fundamental Z$_2$ charge, the liquid state can avoid the condensation of disclinations only by forming composite quasiparticles from Z$_2$ charge attached to disclinations. Such quasiparticles are fractionalized.

The physics behind this is very similar to that of fractional quantum Hall liquids. The Berry curvature of the melted disclination lattice shows up in the effective field theory (\ref{FieldTheory3}) as a topological Lagrangian density term $\mathcal{L}_{t} \sim (\nu/2\phi_0)\, j_\mu \mathcal{J}_\mu$, where $j_\mu$ is the fundamental Z$_2$ charge current (\ref{Z2current}) and $\mathcal{J}_\mu = \epsilon_{\mu\nu\lambda}\partial_\nu\sigma_\lambda$ is the disclination current of the $\phi_0=2$ flux quanta \cite{Nikolic2019}. Note that $\nu=2$ is nominally the simplest unfractionalized state with a Berry curvature because only elementary $J_\mu=2j_\mu$ microscopic excitations, which can be created by local operators, can acquire a conventional $2\pi$ Aharonov-Bohm phase from traversing a path around a flux quantum. These microscopic excitations are bosonic Higgs modes of the nematic fluid; their Z$_2$ charge is $2=0\;(\textrm{mod}\;2)$ in our units, and it takes fractionalization, e.g. $2=1+1\;(\textrm{mod}\;2)$, to obtain quasiparticles with non-trivial Z$_2$ charge that register the flux quanta. 

The topological Lagrangian density must be invariant under all gauge transformations, and it is also desirable to respect the time reversal symmetry. The level-1 Chern-Simons form $\mathcal{L}_{t} \sim (c\nu/2\phi_0)\, \epsilon_{\mu\nu\lambda} \sigma_\mu \partial_\nu \sigma_\lambda$ is not invariant under large gauge transformations. This rules out topological orders that break the time reversal symmetry; we have to construct at least a level-2 theory. Let us introduce two auxiliary gauge fields $\sigma_{\mu}^{+}$ and $\sigma_{\mu}^{-}$, such that the oriented strings of $\sigma_{\mu}^{\pm}$ are outgoing from the $N=\pm1$ singularities and incoming to the $N=\mp1$ singularities respectively. We assume that at least one of $\sigma_{\mu}^{\pm}$ is zero at every point in space, i.e. the strings never intersect. The large gauge transformations exchange the auxiliary gauge fields $\sigma_{\mu}^{\pm}({\bf x},\tau) \to \sigma_{\mu}^{\mp}({\bf x},\tau)$ on individual strings, and the original Z$_{2}$ gauge field is given by
\begin{equation}
\sigma_{\mu}^{\phantom{x}}=\sigma_{\mu}^{+}-\sigma_{\mu}^{-} \ .
\end{equation}
The topological Lagrangian density
\begin{equation}\label{Top}
\mathcal{L}_{t}^{\phantom{x}}=-\frac{\kappa|\hat{{\bf d}}|^{2}c}{2\phi_{0}}\epsilon_{\mu\nu\lambda}^{\phantom{x}}\left(\begin{array}{cc}
\sigma_{\mu}^{+} & \sigma_{\mu}^{-}\end{array}\right)\left(\begin{array}{cc}
\nu^{++} & \nu^{+-}\\
\nu^{-+} & \nu^{--}
\end{array}\right)\partial_{\nu}^{\phantom{x}}\left(\begin{array}{c}
\sigma_{\lambda}^{+}\\
\sigma_{\lambda}^{-}
\end{array}\right)
\end{equation}
is invariant under time reversal $\sigma_{i}^{\pm}({\bf x},\tau) \to \sigma_{i}^{\mp}({\bf x},-\tau)$, $\sigma_{0}^{\pm}({\bf x},\tau) \to -\sigma_{0}^{\mp}({\bf x},-\tau)$ if
\begin{equation}
\nu^{++}=-\nu^{--}\equiv\nu\quad,\quad\nu^{+-}=-\nu^{-+}\equiv\nu' \ .
\end{equation}
The invariance under large gauge transformations then follows due to the presence of the charge factor $c$ in (\ref{Top}). If independent variations of $\sigma_{\mu}^{\pm}$ were allowed, then the combination of (\ref{Top}) and (\ref{FieldTheory3}) would yield the following field equations
\begin{equation}
j_{\mu}^{\phantom{x}}\sim\frac{\nu}{\phi_{0}}\epsilon_{\mu\nu\lambda}^{\phantom{x}}\partial_{\nu}^{\phantom{x}}\sigma_{\lambda}^{+}\quad,\quad j_{\mu}^{\phantom{x}}\sim\frac{\nu}{\phi_{0}}\epsilon_{\mu\nu\lambda}^{\phantom{x}}\partial_{\nu}^{\phantom{x}}\sigma_{\lambda}^{-} \ .
\end{equation}
However, only one of $\sigma_{\mu}^{\pm}$ can be varied independently in any region of space-time while the other is kept at zero. A symmetrized and gauge-invariant formulation of the field equations admits two currents:
\begin{eqnarray}
j_{\mu}^{\phantom{x}} &=& \frac{\nu c}{\phi_{0}}\epsilon_{\mu\nu\lambda}^{\phantom{x}}\partial_{\nu}^{\phantom{x}}(\sigma_{\lambda}^{+}-\sigma_{\lambda}^{-}) \\
\widetilde{j}_{\mu} &=& \frac{\nu}{2\phi_{0}}\epsilon_{\mu\nu\lambda}^{\phantom{x}}\partial_{\nu}^{\phantom{x}}(\sigma_{\lambda}^{+}+\sigma_{\lambda}^{-}) \ . \nonumber
\end{eqnarray}
According to their transformations under time reversal, $j_\mu$ is a fundamental Z$_2$ charge current and $\widetilde{j}_\mu$ is a spin current. The first equation predicts that $\nu$ units of fractional Z$_{2}$ charge are attached to all disclination singularities. The filling factor must be rationally quantized in incompressible quantum liquids as $\nu=p/q$, where $p,q$ are mutually prime integers. Writing $\nu=2p/2q$ and recalling that the nominal Z$_2$ charge of microscopic excitations is 2, we find that the fractionalization of bosons admits any $q\in\mathbb{Z}$, giving an even denominator. Furthermore, the fractional quasiparticles carry another quantum number that transforms non-trivially under time reversal, like spin. A factor of $1/2$ is included in the definition of $\widetilde{j}_\mu$ in order to ensure that $\uparrow$ and $\downarrow$ states are not equivalent in the Z$_2$ algebra.

The large gauge symmetry implements the Z$_2$ algebra and restricts the fusion rules for the fractional quasiparticles. Consider the simplest Laughlin-type states with $\nu=1/q$. A quasiparticle that binds $n_{q}/q$ units of Z$_{2}$ charge to $n_{m}$ disclinations can be labeled as $(n_{q},n_{m})$ in the $c=+1$ gauge, where $n_{q},n_{m}\in\mathbb{Z}$. A bound state of $2q$ elementary fractional quasiparticles is a microscopic excitation
\begin{equation}
(1,n_{m})\times\cdots\times(1,n_{m})\equiv(2q,2qn_{m})\equiv(0,0) \ .
\end{equation}
In these units, $n_{q}=q$ is the fundamental Z$_{2}$ charge quantum, so $n_{q}=2q$ is indistinguishable from a neutral state by $2=0\;(\textrm{mod}\;2)$. Similarly, $2n_{m}$ is the neutral topological charge. These restrictions yield the following fusion rule:
\begin{eqnarray}\label{Fusion}
&& (n_{q1},n_{m1})\times(n_{q2},n_{m2}) = \\
&& \qquad = \bigl(n_{q1}+n_{q2}\;(\textrm{mod}\;2q),\;n_{m1}+n_{m2}\;(\textrm{mod}\;2)\bigr) \ . \nonumber
\end{eqnarray}
We see that $(n_q,1)$ and $(2q-n_q,1)$ are antiparticles to each other.

The gauge-invariant topological Lagrangian density (\ref{Top}) depicts two species ($\pm$) of fractional quasiparticles with a non-trivial exchange statistics. The $\pm$ quasiparticle $(n_{q},n_{m})_{\pm}$ carries fractional charge $n_{q}^{\phantom{x}}e/q$ and the flux $n_{m}^{\phantom{x}}m$ of the $\sigma_{\mu}^{\pm}$ gauge field. The statistical phase for the exchange of two identical quasiparticles $(n_{q},n_{m})_{s}$ is proportional to their fundamental and topological charges
\begin{equation}
\Phi_{c;s,n_{q},n_{m}}=em\,s\nu\,cn_{q}n_{m}=\pi\,s\nu\,cn_{q}n_{m} \ ,
\end{equation}
where we assumed a concrete handedness of adiabatic braiding. We must get a trivial phase $2\pi$ if the quasiparticles are not fractionalized, $n_{q}\;(\textrm{mod}\;2q)=0$, so $em=\pi$ (note that we would get the ``usual'' vortex-like $em=2\pi$ if the charge unit $1$ were assigned to the microscopic excitation instead of its $1/2$ fraction). Since the two fractional quasiparticles are connected by a string, a large gauge transformation inside a region whose boundary does not intersect the string must affect both quasiparticles. The ensuing exchange of $\sigma_{\mu}^{\pm}$ fluxes amounts to $s\to-s$ (effectively $\nu\to-\nu$) without changing the sign of $n_{m}$, while the concurrent sign change of $c$ keeps the phase gauge invariant. This should be interpreted as $\Phi_{c;s,n_{q},n_{m}}=\Phi_{-c;-s,n_{q},n_{m}}$ under gauge transformations. Likewise, time reversal symmetry is respected because $\Phi_{c;s,n_{q},n_{m}}=-\Phi_{c;-s,n_{q},n_{m}}$. The sign factor acquired in the time reversal can be attributed to the reversal of the adiabatic exchange handedness, or equivalently to $s\to-s$ by exchanging the $\sigma_{\mu}^{\pm}$ fluxes. The exchange of different quasiparticles $(n_{q},n_{m})_{s}$ and $(n_{q},n_{m})_{-s}$, picks the phase
\begin{equation}
\Phi_{c;s,-s;n_{q},n_{m}}=\pi\,s\nu'cn_{q}n_{m} \ .
\end{equation}
The gauge symmetry and time reversal are respected, but the rotational symmetry requires $\Phi_{c;+-}=\Phi_{c;-+}$ and hence $\nu'=0$ (in fact, $\Phi_{c;+-}=\Phi_{c;-+}$ must hold in rotations by an arbitrary angle). Overall, the elementary composites of the $\nu=1/q$ Laughlin-like states can be labeled by $(n_{q},1)_\pm$ and $(n_{q},0)_\pm$, where $n_{q}$ is defined $(\textrm{mod}\; 2q)$ and $\pm$ refers to spin. The non-trivial exchange statistics for identical particles of equal spin is
\begin{equation}
(n,1)\leftrightarrow(n,1) = \frac{\pi}{q}n \ ,
\end{equation}
and the full monodromy for distinct quasiparticles of equal spin, under the loop braiding or two exchanges that restore the original state, is
\begin{eqnarray}
(n_{1},1)\leftrightarrow(n_{2},1) &=& \frac{\pi}{q}(n_{1}+n_{2}) \ , \\
(n,0)\leftrightarrow(n',1) &=& \frac{\pi}{q}n \ . \nonumber
\end{eqnarray}
Two particles with different spins have a trivial braiding statistics due to time-reversal and rotational symmetries. 

The elementary quasiparticles and lowest-energy excitations are $(\pm1,1)$ with the self statistical phase $\pm \pi/q$. The Z$_{2}$ structure and the microscopic bosonic statistics constrains $q$ to integers in our units. The case $q=1$ makes the elementary particles fermions, denoted as $\varepsilon = e\times m$ in the context of the toric code model \cite{Kitaev2003, Kitaev2006b}. Here $\varepsilon$ carries spin, in contrast to the $\varepsilon$ of the toric code, but the spin does not define a separate topological sector (at best, a symmetry is needed for its conservation). The $q=1$ state has a topological order analogous to that of the gapped RVB spin liquid envisioned in frustrated magnets, with a possibly different mutual statistics of different-spin particles. Note, however, that the typical fermionic statistics of spinons in RVB spin liquids is imposed by the slave boson theory construction; unlike the Chern-Simons theory, the slave-boson method cannot by itself determine the statistics in two dimensions. A clear thermodynamic distinction of the $q=1$ state from the gapped spin liquid is found only at the system boundary where the flux conservation and the Berry curvature (\ref{Top}) suppress the back-scattering of spin currents and support gapless edge states, see Appendix \ref{appEdge}. In this sense, the $q=1$ nematic and gapped spin liquids have the same relationship as the topologically non-trivial and trivial phases respectively of quantum spin-Hall systems -- but, with an underlying topological order which cannot occur in the weakly interacting spin-Hall materials. The case $q=2$ describes semion quasiparticles $(+1,1)$ and their antiparticles $(-1,1)$ whose exchange statistics has a spin-Hall-like handedness (invariant under time-reversal but opposite for particles and antiparticles). We expect that stable incompressible liquids cannot be characterized by larger values of $q$ because their ground state degeneracy $2q$ would have to be larger than the maximum given by the four classical topological sectors on a torus. Hierarchical states with two fermion flavors seem to be possible.

The ground-state degeneracy on topological manifolds is directly related to the fractional exchange statistics. The ground state degeneracy can be determined by considering how vacuum instantons lift the degeneracy of classical field configurations in different topological sectors. If an $n\ge2$ spin-nematic is placed on a torus, then there are four classical topological sectors: a disclination may be absent or threaded through each of the torus openings. Visualizing the torus in a three-dimensional embedding space, the passage of a disclination through a torus opening corresponds to a parallel non-contractible Dirac-string loop lying on the torus. It turns out that the theory (\ref{FieldTheory3}) assigns different energies to the classical ground states in the four topological sectors. The lowest energy state is a uniform nematic order. If a single non-contractible Dirac loop traverses the torus, then the sign-change $\hat{\bf d}\to-\hat{\bf d}$ across the loop requires gradual variations of $\hat{\bf d}$ over the rest of the torus manifold; the minimum energy cost $E_{\textrm{min}}$ is arranged with $|\partial_\mu\hat{\bf d}| \sim 1/L$, where $L$ is the linear torus size, and then $E_{\textrm{min}}\propto\int d^2x (\partial_\mu\hat{\bf d})^2 \sim \mathcal{O}(L^0)$ is finite. Naively, vacuum instantons have no classical degeneracy to lift.

However, the above picture is incomplete. Gauging a field theory is the means to capture the dynamics of topological defects even when the long range order is lost. But, the Z$_2$ gauge field cannot describe skyrmions and other topological structures that can exist in nematic fluids. This calls for a non-Abelian gauge field $A_\mu^a$ in an extended theory
\begin{equation}
\mathcal{L}=\frac{\kappa}{2}(\partial_{\mu}^{\phantom{x}}\hat{d}^{a}+c\sigma_{\mu}^{\phantom{x}}\hat{d}^{a}+a_0^{\phantom{x}}\delta_{\mu,0}^{\phantom{x}}\hat{d}^a+A_{\mu}^{a})^{2}+\frac{\lambda}{4}F_{\mu\nu}^{a}F_{\mu\nu}^{a}+\cdots
\end{equation}
where
\begin{equation}
F_{\mu\nu}^{a}=\partial_{\mu}^{\phantom{x}}A_{\nu}^{a}-\partial_{\nu}^{\phantom{x}}A_{\mu}^{a}+f^{abc}A_{\mu}^{b}A_{\nu}^{c}
\end{equation}
is the non-Abelian field tensor. The non-Abelian gauge field easily neutralizes the energy of $\hat{d}^a$ variations. For example, a disclination threaded through the $y$ opening of the torus
\begin{equation}
\theta=\frac{2\pi x}{L}\quad,\quad\hat{d}^{a}=\delta^{a,1}\cos\left(\frac{\theta}{2}\right)+\delta^{a,2}\sin\left(\frac{\theta}{2}\right)
\end{equation}
is screened by
\begin{equation}
A_{\mu}^{a}=-\partial_{\mu}^{\phantom{x}}\hat{d}^{a}=\delta_{\mu,x}\frac{\pi}{L}\left\lbrack \delta^{a,1}\sin\left(\frac{\theta}{2}\right)-\delta^{a,2}\cos\left(\frac{\theta}{2}\right)\right\rbrack
\end{equation}
without costing any non-Abelian Maxwell energy ($\lambda$). With this patch in place, all classical topological sectors become degenerate in the thermodynamic limit, even if we take the $\lambda\to\infty$ limit (after $L\to\infty$). A large $\lambda$ is the regime of suppressed skyrmions and suppressed emergent non-Abelian gauge bosons that we are interested in. Note that the introduced non-Abelian gauge symmetry affords the compensation of $\hat{d}^a$ variations even when nematic long-range order (global broken symmetry) is absent. A separate topological order involving nematic hedgehogs may be able to exist in three spatial dimensions \cite{Nikolic2019} through the fluctuations of $A_\mu^a$, but this is not a target for the present analysis.

We show in Appendix \ref{appVacInst} that vacuum instantons of a spin nematic cannot change the spectrum of the classical vacua in separate topological sectors. Therefore, incompressible nematic quantum liquids possess four degenerate ground states on the torus. This admits fermionic and semionic exchange statistics of fractional quasiparticles. Nematic quantum liquids carry a non-trivial topological invariant in the bulk, a fractionalized Z$_2$ ``Chern number''. The system boundary unavoidably hosts topologically protected gapless edge states with a linear energy dispersion, as shown in Appendix \ref{appEdge}. This is phenomenologically similar to a spin-Hall effect, and possibly provides the best chance to experimentally detect a nematic incompressible quantum liquid (which would be necessarily fractional by the virtue of being bosonic).

\section{Discussion and conclusions}\label{secConclusions}

We constructed a continuum-limit Z$_2$ gauge theory of quantum nematic fluids in two spatial dimensions. The dynamics of disclinations is captured by a Z$_2$ gauge field, but the price of the continuum-limit description is that the gauge field must be a vector as if it were a U(1) gauge field. The theory still has only a Z$_2$ local symmetry, but admits in addition a set of large gauge transformations which identify both vortices and antivortices of the vector gauge field with the physical Z$_2$ disclinations. The formal structure of the theory admits the analysis of fractionalized phases which parallels that of the fractional quantum Hall liquids. However, the large gauge transformations render certain features physically equivalent. With this notion, we derived the fusion rules for fractional quasiparticles and explored their exchange statistics. The quasiparticles have either fermionic or semionic statistics, consistent with the topologically degenerate ground states on the torus. The latter is confirmed by considering the effect of vacuum instantons within the Hilbert space of classically distinct topological sectors.

The discovered nematic fractionalized state with fermionic quasiparticles has the same symmetry and equivalent topological order as the gapped Z$_2$ spin liquid, judging by the fusion and braiding rules. However, it is thermodynamically distinct from the gapped spin liquid by the presence of gapless edge states. This is analogous to the distinction between the topologically non-trivial and trivial insulating phases of quantum spin-Hall systems. Therefore, nematic fluids provide a rare possibility of a spin-Hall-like ``topological phase transition'' between states that have the same but non-trivial topological order.

Even though spin nematics can be obtained from physical $S\ge 1$ local moments that exist in many magnetic materials, it is not easy to envision an experimental observation of the predicted fractionalized phases. However, the plausible host models presented in Appendix \ref{appModel} can perhaps be analyzed numerically, e.g. with exact diagonalization on small lattices with $\sim 30$ sites. The microscopic degrees of freedom are spins with minimally three basis states on each site, so there is also room for implementing the models in cold atom gases and quantum simulators.

Our findings provide an entry in the topological classification of incompressible quantum liquids based upon homotopy groups: the relevant homotopy group for nematic fluids is $\pi_1(\mathbb{R}P^n)=\textrm{Z}_2$ if $n\ge2$. The character of these quantum fluids is similar to that of fractional quantum Hall states, but still qualitatively different: the fusion rules and possibilities for fractional statistics are reduced. A broader significance is tied to the peculiar aspect of dynamics visible in the formalism. The Z$_2$ topological defects are treated as if they were vortices whose topological charge matters only modulo 2. This also means that vortices and antivortices are identical. The dynamics is invariant under time reversal, and an Abrikosov lattice of vortices is equivalent to a natural lattice of vortices and antivortices, or visons, through a large gauge transformation. Peculiar arrays of defects and antidefects have been found in various studies of systems with spin-orbit coupling \cite{Nikolic2011a, Nikolic2014, Nikolic2014a, Nikolic2019b}, and the question is whether fractionalized quantum liquids can be stabilized by quantum melting of such defect lattices. The present analysis offers a glimpse to the answer, and suggests the nature of fractionalization that could be anticipated in these cases.

\section{Acknowledgements}\label{secAck}

This research was partly supported by the Department of Energy, Basic Energy Sciences, Materials Sciences and Engineering Award DE-SC0019331. Support was also provided by the Quantum Science and Engineering Center at George Mason University. \\

\appendix

\section{Microscopic model Hamiltonians}\label{appModel}

The nematic tensor field $D^{ab}$ can be derived from spin degrees of freedom as
\begin{equation}
D^{ab} = \frac{1}{2} \lbrace S^a, S^b \rbrace \ ,
\end{equation}
where $S^a$ are spin projection operators in representation $S$, and the braces indicate an anticommutator. This is a symmetric Hermitian operator by construction, but non-trivial only in $S\ge1$ representations. A simple nematic model can then be given by the following Hamiltonian in which a nematic tensor degree of freedom is placed on every lattice site $i$,
\begin{equation}
H = \sum_i \left\lbrack -t\sum_{j\in i} D_i^{ab} D_j^{ab} -u D_i^{ab}D_i^{ab} \right\rbrack \ .
\end{equation}
This model and its short length-scale variations have the continuum limit (\ref{FieldTheory1}) without a background Z$_2$ charge. Constructing a basic nematic lattice model in terms of the director field $\hat{d}^a\sim S^a$ is equally straight-forward:
\begin{equation}\label{LatticeTheory2}
H = \sum_i \left\lbrack -h\sigma^x_{ij} -J\sum_{j\in i} S_i^a \sigma^z_{ij} S_j^a \right\rbrack - \gamma \sum_p \prod_{ij}^{\square_p} \sigma^z_{ij} \ .
\end{equation}
This is, however, a Z$_2$ gauge theory with more degrees of freedom than allowed microscopically ($\sigma_{ij}^z$, $\sigma_{ij}^x$ are Pauli matrices for the quantum Z$_2$ gauge field). On the square lattice, the Z$_2$ gauge transformations generated by local operators
\begin{equation}\label{GaugeGenerator}
G_i = \prod_{j\in i} \sigma^x_{ij} \times \mathcal{T}_i^{\phantom{x}}
\end{equation}
eliminate a set of redundant gauge degrees of freedom. Here, $\mathcal{T}_i^{\phantom{x}}$ is the operator which executes the time reversal of the spin $S_i^a \to -S_i^a$ on the site $i$. Projecting back to the physical Hilbert space amounts to imposing a gauge-invariant local constraint
\begin{equation}\label{Constr}
G_i = (-1)^{2S} \ .
\end{equation}
This is required by the time reversal transformation
\begin{equation}
\hat{d}^a S^a|\hat{\bf d}\rangle=S|\hat{\bf d}\rangle \quad\Rightarrow\quad \mathcal{T}|\hat{\bf d}\rangle = (-1)^{2S}|-\hat{\bf d}\rangle
\end{equation}
of the spin coherent states $|\hat{\bf d}\rangle$ which represent the director field. As a consequence, the Z$_2$ gauge theory will be odd (even) for odd-integer-half (integer) spins. A different ``odd/even'' pattern has been proposed in Ref.\cite{Grover2007} by associating $\hat{\bf d}$ to the direction perpendicular to spin ${\bf S}$, which consistently works only for integer spins $S$ and yields an ``odd/even'' theory for odd/even $S$. The present association of $\hat{\bf d}$ with the spin direction is more general by including non-integer spin magnitudes, and easier to implement in a model.

The Z$_2$ gauge theory (\ref{LatticeTheory2}) has an advantage that it allows a transparent inclusion of the other model ingredients which are needed for stabilizing the incompressible quantum liquids. First of all, we can pick $\gamma<0$ in order to stimulate the appearance of disclinations, or visons in the Z$_2$ gauge theory context, on plaquettes $p$:
\begin{equation}\label{DisclFlux}
\phi_p = \prod_{ij}^{\square_p} \sigma^z_{ij} \to -1 \ .
\end{equation}
Having only the $\gamma<0$ coupling aims to place a vison on every lattice plaquette. Some short-range repulsive interactions $v_{pq}>0$ are needed to space-out the visons,
\begin{equation}\label{LatticeTheory2b}
H_v = \sum_{p\neq q} v_{pq} (1-\phi_p)(1-\phi_q) \ .
\end{equation}
This vison-vison repulsion needs to be strong enough relative to the exchange $J$, otherwise the director stiffness will bring visons into small dipoles. If all conditions are right, the ordered nematic ground state can host a lattice of localized spaced-out visons. This is the parent state of the incompressible quantum liquids we discuss in the main text.

Another important ingredient is the non-zero Z$_2$ charge density which appears in the topological term (\ref{Top}) of the continuum-limit theory. The constraint (\ref{Constr}) is the Gauss' law of the Z$_2$ gauge theory: if $\sigma^z_{ij}$ is the vector potential, then $\sigma^x_{ij}$ is its canonically conjugate electric field operator, and the parity under time reversal of the spin on site $i$ plays the role of the local Z$_2$ charge in (\ref{Constr}). The ``odd'' Z$_2$ gauge theory, therefore, contains one Z$_2$ charge unit on every lattice site, but this is not a dynamical degree of freedom which can form fractional quasiparticles. Introducing a dynamical Z$_2$ charge amounts to doping the spin-nematic model with particles in a different spin representation. There are a couple of ways one might try to achieve this.

One approach is to replace the spin operators in the model (\ref{LatticeTheory2}) with
\begin{equation}
S_i^a = \sum_{r=1,2} \psi_{r,i;\alpha}^\dagger \frac{\mathcal{S}^{a\phantom{\dagger}}_{r;\alpha\beta}}{S_r} \psi_{r,i;\beta}^{\phantom{\dagger}} \quad,\quad \sum_{r=1,2}\psi_{r,i;\alpha}^\dagger\psi_{r,i;\alpha}^{\phantom{\dagger}} = 1 \ ,
\end{equation}
which utilizes slave fermion fields $\psi_{r,i;\alpha}$ in two spin representations
$r=1,2$ of different parities under time reversal; $\mathcal{S}^a_r$ are spin projection matrices in the given representations, and $S_r$ are the corresponding spin magnitudes. One must ensure a single spin on every lattice site by stimulating the presence of slave fermions with chemical potentials $\mu_r>0$ while prohibiting double occupancy with strong repulsive interactions $U,U'>0$:
\begin{eqnarray}\label{LatticeTheory2c}
H_\psi &=& \sum_i \biggl\lbrace \sum_r \biggl\lbrack -\mu_r^{\phantom{\dagger}}\psi_{r,i;\alpha}^\dagger\psi_{r,i;\alpha}^{\phantom{\dagger}} - \omega_r^{\phantom{\dagger}} (\psi_{r,i;\alpha}^\dagger\boldsymbol{\mathcal{S}}^{\phantom{\dagger}}_{r;\alpha\beta}\psi_{r,i;\beta}^{\phantom{\dagger}})^2 \nonumber \\
&& + U (\psi_{r,i;\alpha}^\dagger\psi_{r,i;\alpha}^{\phantom{\dagger}})^2 \biggr\rbrack + U' \psi_{1,i;\alpha}^\dagger\psi_{1,i;\alpha}^{\phantom{\dagger}} \psi_{2,i;\beta}^\dagger\psi_{2,i;\beta}^{\phantom{\dagger}} \biggr\rbrace \nonumber \\
&& -t' \sum_{ij} (\psi_{1,i;\alpha}^{\dagger}\psi_{2,j;\beta}^{\dagger}\sigma^{z\phantom{\dagger}\!}_{ij}\psi_{2,i;\beta}^{\phantom{\dagger}}\psi_{1,j;\alpha}^{\phantom{\dagger}} + h.c.) \ .
\end{eqnarray}
The $\omega_r>0$ terms are needed to push the local spins to their maximum magnitude. The $t'$ term introduces density fluctuations that evenly spread out the slave fermions of each species throughout the lattice, without violating the rotational and time reversal symmetries. Note that direct hopping of single slave fermions is not welcome because it breaks the Z$_2$ symmetry of the nematic fluid. For the same reason, the $t'$ term must be gauged. The gauge transformation generator must also be adjusted as
\begin{equation}
G_i = \prod_{j\in i} \sigma^x_{ij} \times (-1)^{\psi_{2,i;\alpha}^\dagger \psi_{2,i;\alpha}^{\phantom{\dagger}}} \mathcal{T}_i^{\phantom{x}}
\end{equation}
in order to assure commutation with $H_\psi$. Note that only one slave fermion species appears in $G_i$.

The full Z$_2$ Hamiltonian made from (\ref{LatticeTheory2}), (\ref{LatticeTheory2b}) and (\ref{LatticeTheory2c}) is a plausible nematic model which should be able to host fractionalized incompressible quantum liquids of disclinations. Integrating or tracing out the gauge and slave fermion fields would render the same theory in terms of the native nematic tensor operators, but its form introduces additional complexity and we will not pursue it here.

A similar alternative model can be constructed using bosonic auxiliary fields. Let us put on each lattice site a spin degree of freedom in a reducible representation which combines two irreducible representations $S_1$ and $S_2$ of different parities. Let us also introduce another Ising variable $s_i^z$ on every lattice sites which selects one of the two representations via the operator
\begin{equation}
\mathcal{R}_i = \frac{1}{\sqrt{S_1}}\,\mathcal{P}_1\otimes\frac{1+s^z_i}{2} + \frac{1}{\sqrt{S_2}}\,\mathcal{P}_2\otimes\frac{1-s^z_i}{2} \ ,
\end{equation}
where $\mathcal{P}_1$, $\mathcal{P}_2$ are projectors to the given representation's subspace, and $S_1$, $S_2$ are the corresponding spin magnitudes. Now we can replace the spin operators in the model (\ref{LatticeTheory2}) with
\begin{equation}
S_i^a \to \mathcal{R}_i^{\phantom{x}} S_i^a \mathcal{R}_i^{\phantom{x}} \ .
\end{equation}
The addition to the Hamiltonian (\ref{LatticeTheory2})
\begin{equation}\label{LatticeTheory2a}
H_s = -\epsilon \sum_{ij} \sigma_{ij}^z (s_i^x s_j^x + s_i^y s_j^y) -\lambda\sum_i s_i^z
\end{equation}
dopes the model with Z$_2$ charges ($\lambda\neq0$). Note that an XY-like hopping is needed in order to conserve angular momentum: only an exchange of two spins in different representations is allowed. The price to pay is the presence of a new continuous $\textrm{O}(2)/\textrm{Z}_2$ symmetry which can be spontaneously broken in the phase of delocalized Z$_2$ charges. The $s^{x,y}$ variables effectively introduce a separate $n=2$ nematic subsystem whose topological defects are vortices with an integer-valued topological charge. The slave fermion model does not suffer from this surplus physics. We complete this construction with the appropriate modification of the gauge transformation generator
\begin{equation}\label{Constr2}
G_i = \prod_{j\in i} \sigma^x_{ij} \times s_i^z \mathcal{T}_i^{\phantom{x}} \ .
\end{equation}
Note that the entire Hamiltonian, with $H_s$ included, commutes with the adjusted $G_i$.

\section{The relationship between the lattice and continuum-limit Z$_2$ gauge theories}\label{appContLimit}

The gradient term of the continuum limit theory (\ref{FieldTheory3}) is reduced to a lattice form by discretizing the derivative $\partial_\mu f \to f_j-f_i$ on the neighboring sites $i,j$ of the space-time lattice, and interpreting the gauge field $\sigma_\mu \to \sigma_{ij}$ as a variable living on oriented lattice bonds:
\begin{eqnarray}
&& \int d^{2}x(\partial_{\mu}\hat{d}^{a}+\hat{d}^{a}\sigma_{\mu})^{2} = \\
&& \quad = \int d^{2}x(\partial_{\mu}\hat{d}^{a})^{2}+2\int d^{2}x\,\sigma_{\mu}\hat{d}^{a}\partial_{\mu}\hat{d}^{a}+\int d^{2}x(\sigma_{\mu})^{2} \nonumber \\
&& \quad \to \sum_{i}\sum_{j\in i}(1-\hat{d}_{i}^{a}\hat{d}_{j}^{a})+\sum_{i}\sum_{j\in i}\sigma_{ij}\hat{d}_{i}^{a}(\hat{d}_{j}^{a}-\hat{d}_{i}^{a}) \nonumber \\
&& \qquad\quad +\frac{1}{2}\sum_{i}\sum_{j\in i}(\sigma_{ij})^{2} \nonumber \\
&& \quad = \sum_{i}\sum_{j\in i}(1-\sigma_{ij}^{\phantom{x}})(1-\hat{d}_{i}^{a}\hat{d}_{j}^{a})+\frac{1}{2}\sum_{i}\sum_{j\in i}(\sigma_{ij})^{2} \nonumber \ .
\end{eqnarray}
Since the continuum-limit gauge field $\sigma_\mu$ must compensate the sign-change discontinuities of $\hat{d}^a$, which look like $\partial_\mu\hat{d}^a\to-2\hat{d}^a$ on the lattice, its lattice rendition takes inequivalent values $\sigma_{ij}\in\lbrace0,2\rbrace$. Then, we may redefine the lattice gauge field as
\begin{equation}
\widetilde{\sigma}_{ij} = 1-\sigma_{ij}^{\phantom{x}} \in \lbrace \pm 1 \rbrace
\end{equation}
to obtain the gradient form featured in (\ref{LatticeTheory1})
\begin{equation}
(1-\sigma_{ij}^{\phantom{x}})(1-\hat{d}_{i}^{a}\hat{d}_{j}^{a})+\frac{1}{2}(\sigma_{ij})^{2} \to 1-\hat{d}_{i}^{a}\widetilde{\sigma}_{ij}^{\phantom{x}}\hat{d}_{j}^{a} \ .
\end{equation}
The compact Maxwell coupling in terms of $\sigma_\mu$ is equivalent to the product of $\widetilde{\sigma}_{ij}$ on the bonds of a lattice plaquette as written in (\ref{LatticeTheory1}).

The Z$_2$ charge current (\ref{Z2current}) is also readily converted to its lattice form:
\begin{eqnarray}
j_{\mu} &=& \hat{d}^{a}(\partial_{\mu}\hat{d}^{a}+\hat{d}^{a}\sigma_{\mu}) \to
  \hat{d}^{a}_i(\hat{d}^{a}_j-\hat{d}^{a}_i+\hat{d}^{a}_i\sigma_{ij}^{\phantom{x}}) \nonumber \\
&=& \hat{d}^{a}_i(\hat{d}^{a}_j-\hat{d}^{a}_i\widetilde{\sigma}_{ij}^{\phantom{x}}) \to \hat{d}^{a}_i\hat{d}^{a}_j-\widetilde{\sigma}_{ij}^{\phantom{x}} \nonumber \ .
\end{eqnarray}
The background charge in (\ref{Z2current}) show up as an additive constant on the temporal lattice links. Here we see in detail the ambiguous regularization of the singular part of $\hat{d}^a\partial_\mu\hat{d}^a$. Since the current is meant to be invariant under Z$_2$ gauge transformations, we must multiply the above result by $\widetilde{\sigma}_{ij}$,
\begin{equation}
j_{ij}^{\phantom{x}} = \hat{d}^{a}_i\widetilde{\sigma}_{ij}^{\phantom{x}}\hat{d}^{a}_j + \textrm{const.} \ ,
\end{equation}
precisely in the fashion suggested by the gauge-dependent regularization we adopted in the continuum limit. The basic lattice action is then
\begin{equation}
S = -t\sum_{\langle i,j\rangle} j_{ij} + \cdots
  = -t\sum_{\langle i,j\rangle} \hat{d}^a_i\widetilde{\sigma}_{ij}^{\phantom{x}}\hat{d}^a_j + \textrm{const.} + \cdots
\end{equation}
Note that $j_{ij}=j_{ji}$ and $\widetilde{\sigma}_{ij}=\widetilde{\sigma}_{ji}$ have lost their vectorial quality to the compactness and Z$_2$ structure of the theory.

The other terms of the continuum-limit Lagrangian density are less transparent on the lattice. The topological term (\ref{Top}) is fundamentally meaningful only in the continuum limit as a coarse-grained representation of the microscopic Berry curvature in the single-particle spectrum. We also do not have a large gauge symmetry in the lattice Z$_2$ gauge Hamiltonian. However, just the presence of an external ``magnetic field'' $B$ which nucleates a lattice of disclinations provides a Berry curvature to the carriers of Z$_2$ charge when the disclinations delocalize. We do have a ``gauge-invariant'' Maxwell coupling (\ref{MagnField1}) that achieves this purpose in the continuum limit. Discretizing it is straight-forward: choose $\gamma<0$ on spatial plaquettes in (\ref{LatticeTheory1}) to stimulate the appearance of disclinations or visons, and introduce extended-range repulsive interactions between the disclinations to keep them a certain distance apart:
\begin{equation}
S = \cdots -\sum_p\gamma_p\phi_p + \sum_{p\neq q} v_{pq}(1-\phi_p)(1-\phi_q) \ ,
\end{equation}
Here,
\begin{equation}
\phi_p = \prod_{ij}^{\square_p} \widetilde{\sigma}_{ij} \in \lbrace \pm 1 \rbrace
\end{equation}
is the Z$_2$ flux on a lattice plaquette $p$, $\gamma_p<0$ ($\gamma_p>0$) on the spatial (temporal) plaquettes, and $v_{pq}>0$ (on the spatial plaquettes only) decreases to zero or exponentially fast over a finite lattice distance.

\section{Gapless edge states}\label{appEdge}

Chern-Simons couplings capture the chiral response to external perturbations in low-energy effective theories, and generally lead to gapless edge modes when a symmetry is available to protect them. Hence, we expect Kramers pairs of edge states protected by the time-reversal symmetry in fractionalized spin nematics. We will sketch here the derivation of the edge mode dispersion, mainly to illustrate the role played by the mass of the nematic Higgs bosons, which couple to the Z$_2$ gauge field. Our starting point is the topological Lagrangian density (\ref{Top}) with $\nu'=0$ required by the time-reversal and rotational symmetries:
\begin{equation}
\mathcal{L}_{t}^{\phantom{x}}=-\frac{\kappa|\hat{{\bf d}}|^{2}c}{2\phi_{0}}\,\nu\left(\epsilon_{\mu\nu\lambda}^{\phantom{x}}\sigma_{\mu}^{+}\partial_{\nu}^{\phantom{x}}\sigma_{\lambda}^{+}-\epsilon_{\mu\nu\lambda}^{\phantom{x}}\sigma_{\mu}^{-}\partial_{\nu}^{\phantom{x}}\sigma_{\lambda}^{-}\right) \ .
\end{equation}
The Chern-Simons coupling is gauge-invariant in the bulk, but requires an additional matter field $\theta$ for the gauge invariance at the system boundary. A simple integration by parts reveals the boundary terms which restore the gauge invariance
\begin{eqnarray}
&& L_{t}^{\phantom{x}} = \int d^2 x \, \mathcal{L}'_t \\
&& \quad -\frac{\kappa|\hat{{\bf d}}|^{2}c}{2\phi_{0}}\, \nu\oint dx_{\nu}^{\phantom{x}}\,\Bigl\lbrack\epsilon_{\mu\nu\lambda}^{\phantom{x}}(\partial_{\mu}^{\phantom{x}}\theta^{+})\sigma_{\lambda}^{+}-\epsilon_{\mu\nu\lambda}^{\phantom{x}}(\partial_{\mu}^{\phantom{x}}\theta^{-})\sigma_{\lambda}^{-}\Bigr\rbrack \ , \nonumber
\end{eqnarray}
where $dx_{\nu}=dx\,\hat{n}_{\nu}$ and $\hat{n}_{\nu}$ is locally perpendicular to the boundary. We will scrutinize the nature of the edge matter fields $\theta^\pm$ later. For now, let us simplify the problem by focusing on a straight edge along the $x$-direction ($\hat{n}_{\nu}=\delta_{\nu,2}$) and introduce kinetic energy terms for the emerging edge modes, which are inherited from the similar terms in the bulk. The real-time Lagrangian per unit length on the edge is
\begin{eqnarray}
\mathcal{L}^{\textrm{(e)}} &=& \sum_{s=\pm} \biggl\lbrace \frac{\kappa|\hat{{\bf d}}|^{2}}{2}(\partial_{0}^{\phantom{x}}\theta^{s}+\sigma_{0}^{s})^{2}-\frac{\kappa|\hat{{\bf d}}|^{2}}{2}(\partial_{x}^{\phantom{x}}\theta^{s}+\sigma_{x}^{s})^{2} \nonumber \\
&& +\frac{\kappa|\hat{{\bf d}}|^{2}c}{2\phi_{0}} \nu\,s\Bigl\lbrack (\partial_{0}^{\phantom{x}}\theta^{s})\sigma_{x}^{s}-(\partial_{x}^{\phantom{x}}\theta^{s})\sigma_{0}^{s}\Bigr\rbrack\biggr\rbrace +\langle\textrm{Maxwell}\rangle \nonumber
\end{eqnarray}
We may pick the large gauge $\sigma_{\mu}^{\phantom{x}}=c\sigma_{\mu}^{c}$, $\sigma_{\mu}^{-c}=0$ with either $c=\pm1$, keeping with the requirement that one of $\sigma_{\mu}^{\pm}$ be zero. The matter field $\theta^{\pm}$ associated with the vanishing $\sigma_{\mu}^{\pm}$ is also a figment of the large gauge choice, so the edge theory becomes
\begin{eqnarray}
\mathcal{L}^{\textrm{(e)}} &=& \frac{\kappa|\hat{{\bf d}}|^{2}}{2}(\partial_{0}\theta+c\sigma_{0})^{2}-\frac{\kappa|\hat{{\bf d}}|^{2}}{2}(\partial_{x}\theta+c\sigma_{x})^{2} \nonumber \\
&& +\frac{\kappa|\hat{{\bf d}}|^{2}c}{2\phi_{0}}\,\nu\Bigl\lbrack (\partial_{0}\theta)\sigma_{x}-(\partial_{x}\theta)\sigma_{0}\Bigr\rbrack+\cdots
\end{eqnarray}
If the matter field $\theta$ were soft, as in the XY model, then we would be able to represent its normal modes with $\theta(x,t)=kx-\omega t$ and obtain their dispersion by taking the Fourier transform of $\mathcal{L}^{\textrm{(e)}}\to0$:
\begin{eqnarray}
&& (\omega^{2}-k^{2})+(\sigma_{0}^{2}-\sigma_{x}^{2}) \\
&& \qquad -2c(\omega\sigma_{0}+k\sigma_{x})-\frac{c\nu}{\phi_{0}} (\omega\sigma_{x}+k\sigma_{0})=0 \ . \nonumber
\end{eqnarray}
However, this exploits a U(1) symmetry that a nematic system does not have. The coupling of matter to the gauge field retains only the Z$_2$ gauge symmetry. Fortunately, this is easy to take into account. The nematic Higgs modes, which couple to the Z$_2$ gauge field, are massive. We simply need to include a ``bare'' mass $m$ term in the edge mode dispersion in order to capture all special properties of the Higgs modes that remain visible in the continuum limit with existing symmetries:
\begin{eqnarray}
&& (\omega^{2}-k^{2}-m^{2})+(\sigma_{0}^{2}-\sigma_{x}^{2}) \\
&& \qquad -2c(\omega\sigma_{0}+k\sigma_{x})-\frac{c\nu}{\phi_{0}}(\omega\sigma_{x}+k\sigma_{0})=0 \ . \nonumber
\end{eqnarray}
Note that embedding $m$ into any of the gauge field terms would violate gauge invariance. Now we can solve for $\omega(k)$:
\begin{eqnarray}\label{EdgeDispersion1}
&& \omega(k) = \frac{1}{2}\left(2c\sigma_{0}+\frac{c\nu}{\phi_{0}}\sigma_{x}\right) \\
&& \quad\quad \pm\frac{1}{2}
\Biggl\lbrack \left(2k+2c\sigma_{x}+\frac{c\nu}{\phi_{0}}\sigma_{0}\right)^{\!\!2} +4(m^{2}+\sigma_{x}^{2}-\sigma_{0}^{2}) \nonumber \\
&& \qquad\quad -\left(2c\sigma_{x}+\frac{c\nu}{\phi_{0}}\sigma_{0}\right)^{\!\!2}+\left(2c\sigma_{0}+\frac{c\nu}{\phi_{0}}\sigma_{x}\right)^{\!\!2} \Biggr\rbrack^{\frac{1}{2}} \nonumber
\end{eqnarray}
We have assumed that $\sigma_x$ and $\sigma_0$ do not depend on $x$ and $t$, but the ``chemical potential'' $\sigma_0$ gradually changes with the distance $y$ from the edge. At some specific distance,
\begin{equation}
\sigma_{0}=\pm\sqrt{\sigma_{x}^{2}+4m^{2}\left(\frac{\nu}{\phi_{0}}\right)^{-2}}
\end{equation}
is achieved and the last three terms under the square root in (\ref{EdgeDispersion1}) cancel out, leaving
\begin{equation}
\omega(k)=\left(c\sigma_{0}+\frac{c\nu}{2\phi_{0}}\sigma_{x}\right)\pm\left\vert k+\left(c\sigma_{x}+\frac{c\nu}{2\phi_{0}}\sigma_{0}\right)\right\vert \ .
\end{equation}
Formally, there are two branches of dispersion
\begin{eqnarray}
\omega_{+}&=&k+\left(1+\frac{\nu}{2\phi_{0}}\right)c(\sigma_{0}+\sigma_{x}) \\
\omega_{-}&=&-k+\left(1-\frac{\nu}{2\phi_{0}}\right)c(\sigma_{0}-\sigma_{x}) \ , \nonumber
\end{eqnarray}
but only one of them corresponds to a low-energy edge mode (the other one is microscopically justified in lattice models, e.g. as originating from a high-energy Hofstadter band). The low-energy edge spectrum is gapless, but the zero-energy modes $\omega=0$ are shifted to $k\neq0$ due to $m\neq0$. The modes transform under time-reversal $\sigma_{x}\to-\sigma_{x}$, $\sigma_{0}\to\sigma_{0}$, $k\to-k$, $\omega\to\omega$ into:
\begin{eqnarray}
\omega_{+}&=&-k+\left(1+\frac{\nu}{2\phi_{0}}\right)c(\sigma_{0}-\sigma_{x}) \\
\omega_{-}&=& k+\left(1-\frac{\nu}{2\phi_{0}}\right)c(\sigma_{0}+\sigma_{x}) \nonumber \ .
\end{eqnarray}
We see that both left- and right-propagating modes exist and have the same energy. The time-reversal operation maps a gapless edge mode to its Kramers partner, according to the time-reversal symmetry of the theory. Note that the time-reversal exchange $\sigma_{i}^{\pm}\to\sigma_{i}^{\mp}$, $\sigma_{0}^{\pm}\to-\sigma_{0}^{\mp}$, $\theta^{\pm}\to\theta^{\mp}$ of the $\pm$ effective fields is a symmetry transformation of physical degrees of freedom because the ``large'' gauge field $c$ is kept fixed.

\section{Vacuum instantons}\label{appVacInst}

Here we calculate the spectrum of vacuum instantons in a Z$_2$ gauge theory on a torus. We assume that the four classical topological sectors, corresponding to $0$ or $1$ disclinations threaded through a torus opening, are exactly degenerate. A vacuum instanton is then a quantum tunneling event in which the disclination charge $N_i\in\lbrace 0,1 \rbrace$ through a torus opening $i=1,2$ changes.

The simplest way forward is to examine action costs of instantons on the lattice and then take the continuum limit. The state $N_i=0$ is trivial ($\widetilde{\sigma}_{ij}=+1$ on all lattice links), while the only other option $N_i=1$ corresponds to $\widetilde{\sigma}_{ij}=-1$ placed on a ``ladder'' which extends in the direction $i$. The ladder can be positioned anywhere on the torus and wraps completely around the torus without producing a vison $\Phi=-1$ on any lattice plaquette. A change of $N_i\to0$ at some instant of time $\tau$ corresponds to a disappearance of the ``ladder'', which in turn generates the ``electric'' Z$_2$ flux on the ``stack'' of temporal plaquettes adjacent to the ``ladder'' at time $\tau$. This is the Z$_2$ Faraday effect. Only the ``electric'' flux contributes to the instanton action in proportion to the length of the ``ladder'', i.e. the linear torus size. Conversion to the continuum limit makes this cost positive, $S_i \propto (L/\delta a) \phi_0^2 (\partial_0 N_i)^2$, where $L$ is the system size, $\delta a$ is the ultra-violet cut-off length (lattice constant), and $\phi_0=2$ is the Z$_2$ flux quantum in the appropriate units.

The topological term (\ref{Top}) is also sensitive to instantons, but only at the crossing point of two orthogonal ladders for $N_1=1$ and $N_2=1$ because the Faraday flux at a temporal plaquette due to $\partial_0 N_i$ must pick a non-trivial gauge field on the orthogonal spatial links imposed by $\epsilon_{ij}N_j=1$. The Chern-Simons coupling is, actually, not well-defined on a lattice, given that multiple links are orthogonal to a particular nearby plaquette. Nevertheless, if $N_1$ and $N_2$ change at different instants of time, then the sensible average of $\widetilde{\sigma}_\mu$ on all orthogonal links amounts to simply associating $\sigma_{ij}=2$ ($\widetilde{\sigma}_{ij}=-1$) on one orthogonal link to one temporal plaquette at the crossing point of two ``ladders''. Picking a fixed gauge, this gives a Chern-Simons Lagrangian proportional to $2\phi_0 (N_1\partial_0N_2-N_2\partial_0N_1)$.

Thus, we obtain the following continuum-limit Lagrangian of vacuum instantons
\begin{eqnarray}
L_i &=& \frac{\phi_0^2}{2e^{2}} \frac{L}{\delta a} \Bigl\lbrack(\partial_{0}N_{1})^{2}+(\partial_{0}N_{2})^{2} \Bigr\rbrack \\
&& + \alpha\,(N_{1}\partial_{0}N_{2}-N_{2}\partial_{0}N_{1}) \ . \nonumber
\end{eqnarray}
We assumed that the gradient energy is completely compensated. After introducing the canonical momenta
\begin{eqnarray}
P_{1} &=& \frac{\delta L_i}{\delta\partial_{0}N_{1}} = \frac{\phi_0^2}{e^{2}} \frac{L}{\delta a}(\partial_{0}N_{1}) - \alpha N_{2} \\
P_{2} &=& \frac{\delta L_i}{\delta\partial_{0}N_{2}} = \frac{\phi_0^2}{e^{2}} \frac{L}{\delta a}(\partial_{0}N_{2}) + \alpha N_{1} \ ,
\end{eqnarray}
we obtain the instanton Hamiltonian
\begin{eqnarray}
H_i &=& P_{1}(\partial_{0}N_{1})+P_{2}(\partial_{0}N_{2})-L_i \\
&=& \frac{1}{2}\frac{e^{2}}{\phi_{0}^{2}}\frac{\delta a}{L} \Bigl\lbrack (P_{1}+\alpha N_{2})^{2}+(P_{2}-\alpha N_{1})^{2} \Bigr\rbrack \ . \nonumber
\end{eqnarray}
The key observation here is that the Hamiltonian vanishes in the thermodynamic limit $L/\delta a\to\infty$. The vacuum instanton dynamics is completely suppressed, and the classical ground state degeneracy cannot be lifted.

The above instanton Hamiltonian for nematic fluids is formally similar to that for quantum Hall liquids, but with one crucial difference: the effective mass for instanton fluctuations scales as a system size instead of being finite. This is a direct consequence of how the nematic director field $\hat{d}^a$ and the Z$_2$ gauge fields compensate each other: the Z$_2$ gauge field must be consumed into Dirac strings as long as $\hat{d}^a$ retains coherence across length scales substantially larger than the lattice constant. No such requirements are imposed on the U(1) gauge field in quantum Hall liquids.



%

\end{document}